\documentclass[10pt]{extarticle}
\usepackage{spconf,amsmath,graphicx}
\usepackage{subfiles} % Best loaded last in the preamble
\usepackage{booktabs}
\usepackage{makecell}
\usepackage{multirow}
\usepackage{caption}
\usepackage{subcaption}
 \usepackage{setspace}
\let\OLDthebibliography\thebibliography
\renewcommand\thebibliography[1]{
  \OLDthebibliography{#1}
  \setlength{\parskip}{0pt}
  \setlength{\itemsep}{0pt plus 0.3ex}
}

\newcommand{\modelname}{Papez}
\newcommand{\modelnameUpperCase}{PAPEZ}
\title{\modelnameUpperCase: Resource-Efficient Speech Separation with Auditory Working Memory}
\name{Hyunseok Oh, Juheon Yi, Youngki Lee}
\address{Seoul National University}
\begin{document}
%\ninept
%
\maketitle
\begin{abstract}
Transformer-based models recently reached state-of-the-art single-channel speech separation accuracy; However, their extreme computational load makes it difficult to deploy them in resource-constrained mobile or IoT devices. 
We thus present {\modelname}, a lightweight and computation-efficient single-channel speech separation model.
{\modelname} is based on three key techniques. We first replace the inter-chunk Transformer with small-sized auditory working memory. Second, we adaptively prune the input tokens that do not need further processing. Finally, we reduce the number of parameters through the recurrent transformer.
Our extensive evaluation shows that {\modelname} achieves the best resource and accuracy tradeoffs with a large margin. 
We publicly share our source code at  \texttt{https://github.com/snuhcs/Papez}~.

\end{abstract}

\section{Introduction}
\label{sec:intro}

%Transformer-based models are showing overwhelming performances in the speech separation domain. 

%Resource-efficient speech separation techniques open a way for numerous real-world applications, as mobile or IoT devices are heavily constrained in computational resources. 
Speech separation 
%(separating individual speech source from a mixture of multiple speakers, also known as the cocktail party problem) 
%on mobile/embedded devices
serves as a preparatory stage for various downstream applications, e.g., speech recognition, speaker diarization, and machine translation.
While Transformer-based models have recently achieved state-of-the-art separation performance~\cite{subakan2021sepformer, lutati2022sepit, rixen2022qdpn}, they are severely intensive in computation and memory, making it difficult to deploy them in resource-constrained mobile/IoT devices. 
For example, SepFormer~\cite{subakan2021sepformer} with 26M parameters costs 2.53 sec
%while ConvTasNet\cite{luo2019conv} (5.6M params) requires 251 ms, 
 for inference on an 8kHz, 0.5-sec input with Samsung Galaxy S20 CPU; real-time mobile speech processing is thus infeasible.

% Not revised yet. Planning to write key observations preceding our approaches.
In this paper, we propose {\modelname}, a resource-efficient single-channel speech separation model.
We design {\modelname} with a couple of key observations regarding the inefficiencies in the state-of-the-art transformer-based models.

First, we find that a widely-used dual-path process approach~\cite{subakan2021sepformer,luo2020dualpathrnn} incurs unnecessary computational overhead~\cite{rixen2022qdpn}.
Dual-path process alleviates the excessive processing load of elongated input sequences by chunking the sequence and modeling the intra-chunk and inter-chunk dependency separately~\cite{luo2020dualpathrnn,lam2021sandglasset}.
However, our in-depth analysis finds that only a few transformer layers of the inter-chunk transformer are utilized in effect.
Second, prior models employ a fixed processing path regardless of the input content. 
%Yet, it is much easier to separate some segments of a speech mixture than others, e.g., silence or single-speaker speech~\cite{cosentino2020librimix}. %regardless of the input content. 
%However, the separation difficulty heavily varies depending on the content 
Yet, a meaningful portion of an input signal can be pruned out during the computation since some segments of a speech mixture are much easier to separate than others, e.g., silence or single-speaker speech~\cite{cosentino2020librimix}.
%a mixture composed of overlapping genders 
%with similar timbre and pitch 
%is much more difficult to separate than the one with non-overlapping genders).
%We take two key approaches: (i) design an efficient transformer architecture for speech separation,
%and (ii) design a mechanism that can adjust computation depending on the input content.

Leveraging our observations, 
we develop two key techniques to design {\modelname}:
%we propose {\modelname}, a resource-efficient single-channel speech separation model based on two key techniques:  
\emph{Auditory Working Memory (AWM) Transformer} and  \emph{Adaptive Token Pruning}. 
First, our \emph{AWM Transformer} architecture 
augments the intra-chunk transformer with a small-sized short-term memory to replace the inter-transformer.
%utilizes a small-sized working memory (replacing the inter-chunk transformer) that 
In detail, the short-term memory captures and stores the global context needed for local intra-chunk processing. Note that our architecture is bio-inspired; the human brain has a functionality called AWM that temporarily stores audio features for auditory processing~\cite{pasternak2005working,kumar2016brain}.  
Also, studies in NLP found that global tokens attending  the whole sequence are effective in modeling a very long sequence~\cite{Ainslie2020ETC,Zaheer2020BigBT}.
%Working memory efficiently replaces the 
%The number of added memory tokens can be much smaller than the chunk size, like 16 mem tokens on chunk size of 250. So the inter-chunk processing taking up nearly half the computation time has replaced with the working memory that only adds marginal overhead to the intra-chunk processing. 

Second, our \emph{Adaptive Token Pruning} technique context-adaptively prunes redundant input tokens that need no further processing. In detail, each token self-determines whether to stop or continue processing itself in every transformer layer probabilistically.
%annotates each token with a probability indicating
%the processing termination when it exceeds a certain threshold. The individual sequence token self-determines 
%whether to (i) partially contribute current value to the output, (ii) further process itself, or (iii) stop further computation and commit itself to the output, 
This dynamically scales down the width (the number of tokens) and depth (layer iteration) of the  transformer computation.
%This naturally induces dynamic width and depth scaling of the transformer: the number of on-process tokens (width of the transformer) can be dynamically scaled, and the depth of the transformer stack can be dynamically scaled if no on-process token exists. 
%We also optimize the compute efficiency of probability annotation by piggy-backing the probability estimator to the FFN of the transformer layer and the embedding layer (rather than employing a dedicated estimator as in prior works)
%Prior works introduced a dedicated estimator for the probability annotation, but it incurs additional computational overhead, since each token should feed-forward additional estimator DNN per transformer layer. Thus we piggy-back the probability estimator to the FFN of the transformer layer and the embedding layer. In this way, we do not need a dedicated estimator DNN and it can be computed with existing FFNs simultaneously. 
%The resulting model architecture reduces to a stack of working-memory transformer layers. 
%Lastly, Recurrent Transformer shares weight across multiple transformer layers, further enabling the memory efficiency.
Moreover, we optimize the latency and model size by (i) piggybacking the probability estimator to the feed-forward network (FFN) of the transformer layer (rather than using a dedicated estimator as in prior works~\cite{graves2016adaptive,dehghani2018universal}), and (ii) employing the recurrent transformer architecture that shares weight across transformer layers.

Our extensive evaluation shows that {\modelname} enables a significantly more efficient computation-accuracy tradeoff compared to prior models. Specifically, {\modelname} achieves $3.6\times$, $4.14\times$ smaller parameters and $2.44\times$, $1.67\times$ faster inference latency than Tiny-Sepformer~\cite{luo2022tiny} and A-FRCNN~\cite{hu2021afrcnn} with 4.2dB and 1.1dB higher SI-SNR in WSJ0-2Mix dataset, respectively.
%, and XXdB and XXdB higher SI-SNR than  in Libri2Mix dataset, respectively
 We also reduce the model size of the state-of-the-art Sepformer by $17.7\times$ with minimal accuracy drop. Our techniques are widely applicable on other transformer-based models (e.g., Sepformer variants~\cite{lutati2022sepit,rixen2022qdpn,luo2022tiny}) as well.

\section{Related Work}
\label{sec:relatedworks}

Separating each speech source from a single-channel mixture of multiple speakers 
%(known as the cocktail party problem~\cite{Haykin2005TheCP}) 
is a fundamental yet difficult problem~\cite{lam2021sandglasset}. 
%Single-channel speech separation techniques initially advanced with frequency domain approaches. 
% https://link.springer.com/content/pdf/10.1007/978-3-642-35615-5.pdf
Initially, frequency-domain approaches separate the mixture in its time-frequency STFT representation~\cite{Logeshwari2012ASO}. 

The success of TasNet~\cite{luo2018tasnet} has brought great interest in RNN-based time-domain approaches,  %MSGM-tasnet
which uses the non-STFT encoder to extract an effectively separable representation of the waveform.
The problem is that encoded sequences get unbearably long to model with RNNs since a smaller sliding window of encoder led to higher performance~\cite{luo2020dualpathrnn, chen2020dual,lam2021sandglasset}. 
Dual-path process addresses this by chunking the sequence and modeling intra-chunk and inter-chunk  separately~\cite{luo2020dualpathrnn}. 

Unfortunately, RNN is limited due to its sequential nature; its computation cannot parallelize, and it relies on the inductive bias of temporal invariance~\cite{subakan2021sepformer,lin2021surveyT}. Recently, RNN-free transformer-based models (i.e., Sepformer~\cite{subakan2021sepformer} and its variants~\cite{lutati2022sepit,rixen2022qdpn}) achieved remarkable performance with three factors: (i) maximal connectivity of self-attention mechanism to model long-term dependency, 
 (ii) no inductive bias unlike CNNs (translation invariance) or RNNs (temporal invariance)~\cite{lin2021surveyT}, and (iii) parallelization of computation~\cite{subakan2021sepformer}. Although these models are faster than RNN-Transformer hybrid models~\cite{lam2021sandglasset,chen2020dual}, the dual-path process doubles the time complexity of the Transformer, which is already huge. 
 
A few works focus on reducing the computational load and model size of speech separation models. ConvTasNet~\cite{luo2019conv} replaces LSTM stack of TasNet~\cite{luo2018tasnet} with a stack of 1-D dilated convolutions. A-FRCNN~\cite{hu2021afrcnn} fuses multi-scale features processed with CNN. Yet, CNN is not as effective as Transformer in modeling long-term dependency~\cite{vaswani2017attention}. Tiny-sepformer~\cite{luo2022tiny} cuts down model size with parameter sharing~\cite{wu2020liteT}. However, it suffers from a significant accuracy drop of 4-5dB SI-SNRi in the WSJ0-2Mix dataset~\cite{hershey2016deepCL}.

%\end{document}

\begin{figure}[t]
\centering
\includegraphics[width=0.9\linewidth]{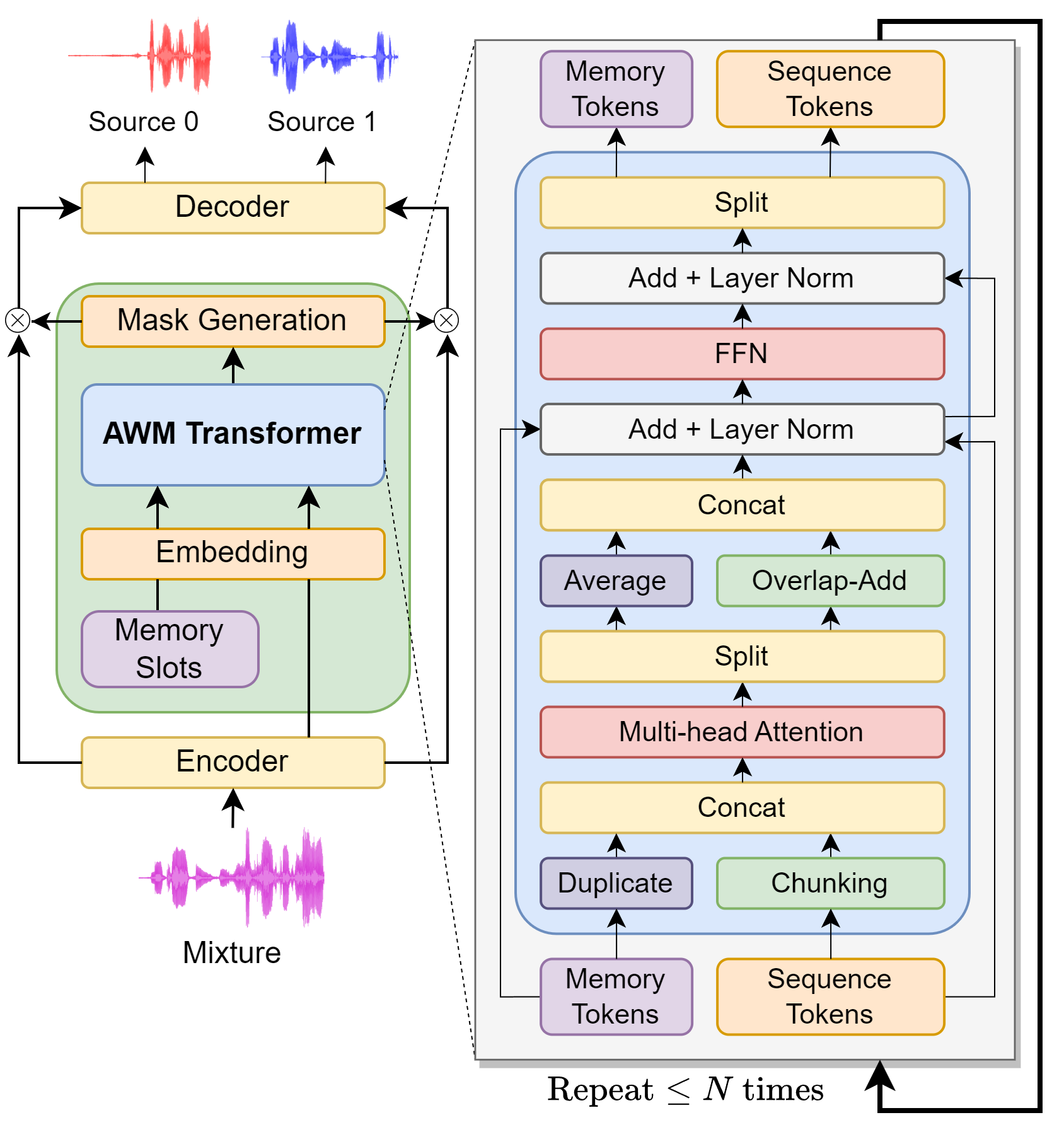}
\vspace{-1em}
\caption{Overall architecture of our \modelname\ model.
The iteration steps of the AWM Transformer layer are determined by our Adaptive Token Pruning technique.}
\label{fig:architecture}
\vspace{-1.75em}
\end{figure}

\section{Approach}
\label{sec:approach}

\subsection{Architecture Overview}
Figure~\ref{fig:architecture} shows the overall model architecture of \modelname. 
%We use the state-of-the-art Sepformer~\cite{subakan2021sepformer} as baseline architecture (other variants~\cite{lutati2022sepit,rixen2022qdpn,luo2022tiny} can also leverage our techniques).
We take the time-domain masking approach~\cite{subakan2021sepformer, luo2019conv} composed of three modules: Encoder, Masking Module, and Decoder. %ConvTasNet
First, the \textit{Encoder} extracts a 2D spectrogram-like representation from the mixture signal. Second, the \textit{Masking Module} estimates the mask over the 2D  representation for each speaker. Finally, the \textit{Decoder} uses the mask and 2D representation to reconstruct the clean speech of each speaker. 

\noindent\textbf{Encoder and Decoder.}
The encoder is a two-layer downsampling  1D convolutional network, which is a sequence of 1-D convolution, instance normalization (IN), ReLU, and point-wise convolution. Similarly, the decoder is a two-layer upsampling convolutional network consisting of point-wise convolution, IN, ReLU, and 1-D transposed convolution. 

\noindent \textbf{Masking Module.} An encoded mixture signal transforms into mask estimates through three stages. First, the \textit{Embedding network} embeds the 2D signal representation into a sequence of tokens. Second, the \textit{AWM transformer} process the token sequence. Finally, the \textit{Mask Generation network} generates masks for each speaker. Embedding and Mask Generation networks are both a two-layer fully-connected feed-forward network with $\text{PReLU}$ activation, and the $\tanh$ activation at the end of the Mask Generation network.

\begin{figure}[t]
\centering
\includegraphics[width=0.95\linewidth]{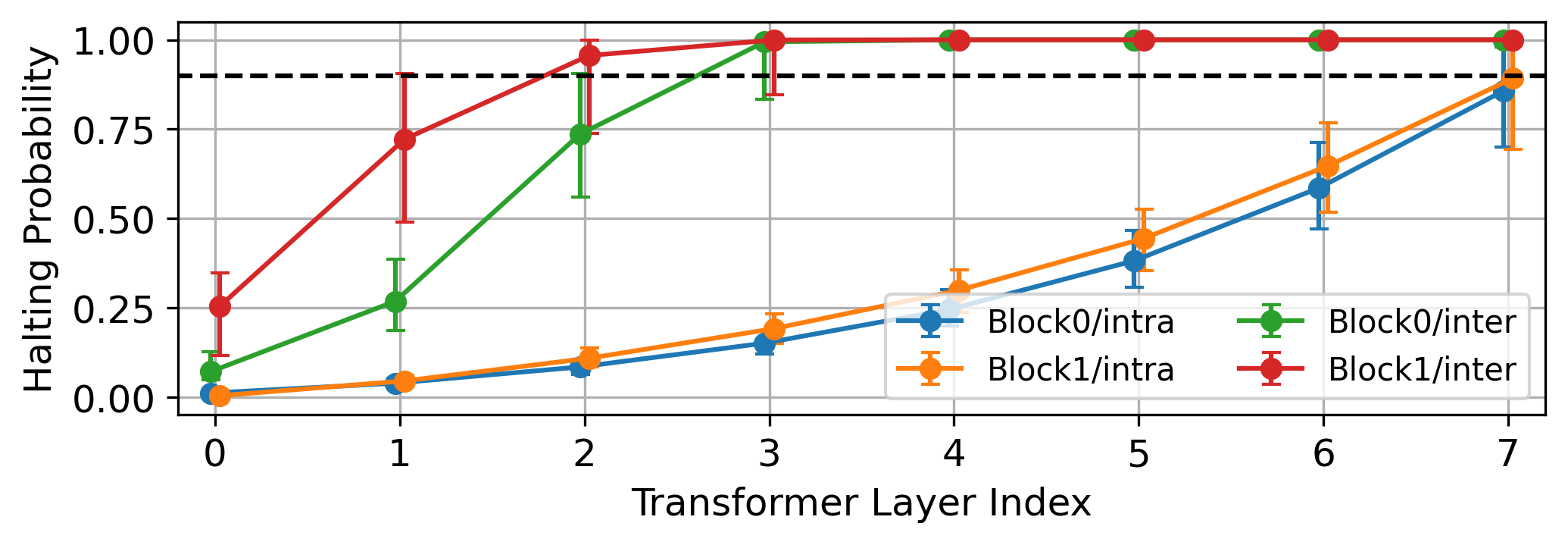}
\vspace{-1em}
\caption{Redundancy of input tokens in the Sepformer's dual-path process. The black dashed line indicates the threshold $P_{th} = 0.9$ of adaptively pruning the redundant token. }
\label{fig:dual_path_analysis}
\vspace{-0.7em}
\end{figure}
\begin{table}[t]
\centering
\fontsize{9}{11}\selectfont
\begin{tabular}{ c | c | c}
  \toprule
  & Time Complexity & Params \\ 
  \toprule
 \makecell{Intra + Inter-T} &   $\mathcal{O}(N\sqrt{N}H^2)$&$8H^2$   \\
 \midrule
 \makecell{Intra +$\frac{1}{S}$Inter-T} & $\mathcal{O}(\frac{N\sqrt{N}}{\sqrt{S}}H^2)$& $(4 + \frac{4}{S})H^2$   \\ %(1+\frac{1}{S})
 \midrule
 \makecell{Intra + AWM} & \makecell{$\mathcal{O}(\frac{N}{K}(K + M)^2H^2)$ \\ $\approx 2NKH^2 $}  & $4H^2$ \\
 \midrule 
 Intra Only & $2NKH^2$ & $4H^2$ \\  
 \bottomrule 
\end{tabular}
\vspace{-0.5em}
\caption{AWM vs. Inter-Transformer. $N$: \# input tokens, $M$: \# memory tokens, $K$: chunk size, $H$: token size, and $S$: depth ratio of intra- and inter-Transformer. $\approx$ when $M \ll K$.}
\label{tab:complexity}
\vspace{-2.25em}
\end{table}

\subsection{Auditory Working Memory (AWM) Transformer}
\label{sec:wmt}

%Sepformer realizes the inter-chunk operation of the dual-path process with the inter-transformer. However, evidences show that this inter-transformer is redundant. 
%Sepformer realizes the inter-chunk operation of the dual-path process with the inter-transformer. However, we find that inter-transformer is redundant.
We find that the inter-chunk transformer of the dual-path process is often redundant 
%QDPN~\cite{rixen2022qdpn} shows it by down-sampling the inter-chunk transformer. Furthermore, 
: as shown in Figure~\ref{fig:dual_path_analysis}, when we apply our Adaptive Token Pruning (Section~\ref{sec:atp}) to the Sepformer, most of the inter-transformer's tokens were quickly pruned after the first 2,3 layers. 
Similar observations were found in \cite{rixen2022qdpn} as well.
%Furthermore, 
However, na\"ively reducing the depth of inter-transformer does not change the asymptotic self-attention complexity $\mathcal{O}(N\sqrt{N})$ for input size $N$ (See  Table~\ref{tab:complexity}.).

In this light, we propose \textit{Auditory Working Memory (AWM) Transformer}, a new transformer architecture augmented with a small-sized short-term memory to replace the compute-intensive inter-transformer. This is realized by special \textit{memory tokens} distinct from input sequence tokens.
%The main difference from the vanilla Transformer~\cite{vaswani2017attention} is how memory tokens are leveraged; (i) 
In detail, the memory tokens are concatenated in front of every sequence chunk to globally attend the input sequence. After the multi-head attention mechanism, the memory tokens of each chunk are averaged to aggregate the global information. Right side of Figure~\ref{fig:architecture} illustrates the detailed mechanism of the AWM transformer layer.  Note that chunking is only applied to multi-head attention, reducing the FFN's latency to nearly $50\%$
%$\approx$0.5$\times$ 
since it halves the number of input tokens.

Our idea is bio-inspired; humans store auditory information like timbre, tone, and pitch~\cite{schulze2013wmpitch}
in a dedicated AWM for up to a few seconds~\cite{pasternak2005working}, which is pivotal for auditory discrimination~\cite{zhang2016auditory}. 
Also, the auditory cortex is interconnected with AWM areas (e.g., hippocampus), and the AWM is consistently maintained and retrieved along with auditory processing~\cite{kumar2016brain}. 
AWM has two advantages over the dual-path process. First, the time complexity is linear to input size, smaller than the transformer-based dual-path process's superlinear complexity as shown in Table~\ref{tab:complexity}. 
Second, it is fully parallelizable with intra-chunk processing. 
%%https://www.ncbi.nlm.nih.gov/pmc/articles/PMC4095971/

\begin{figure}[t]
     \centering
     \includegraphics[width=0.9\linewidth]{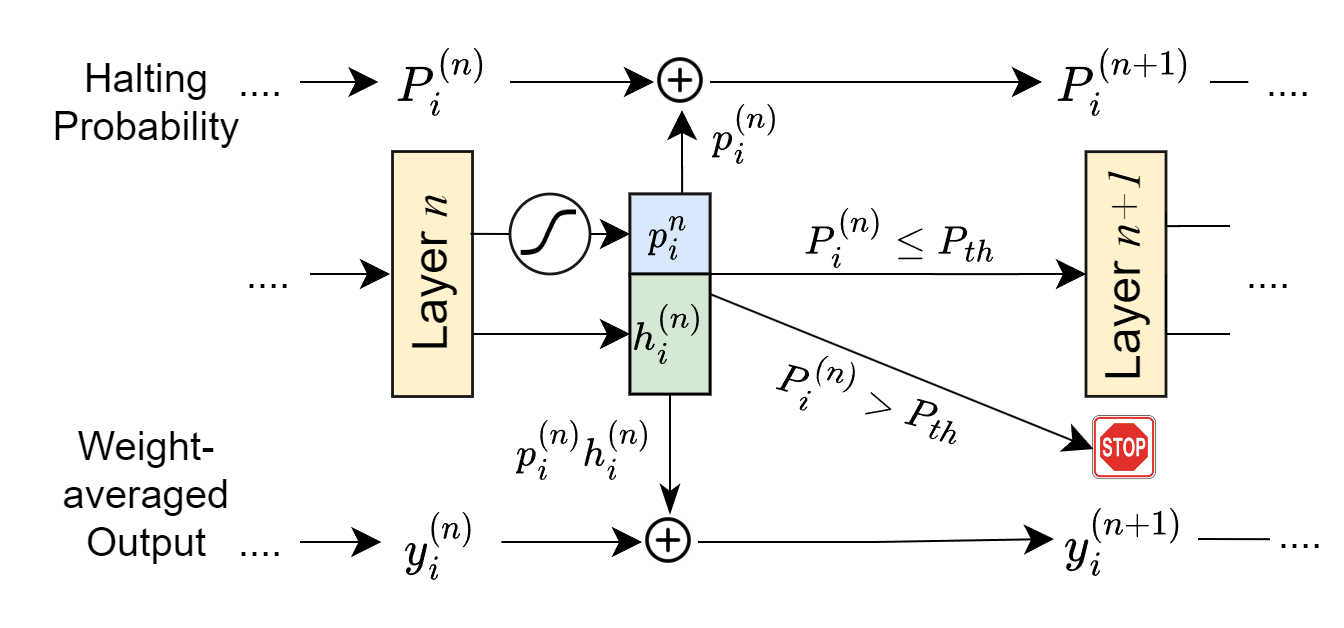}
    \vspace{-1.25em}
     \caption{Operation of Adaptive Token Pruning.}
     \label{fig:adaptivepruning}
\vspace{-2em}
\end{figure}

\subsection{Adaptive Token Pruning}
\label{sec:atp}
In a mixture of speech, some segments are more straightforward to separate than 
%need less "pondering" than 
others, which can be exploited for faster processing. %Adaptive Computation
%The complexity of separating a mixture of speech heavily depends on its content.
For example, silent or non-overlapping speech segments are much easier to separate than overlapped speeches, which consist of $80\%$ of audio inputs in a real-world scenario~\cite{cosentino2020librimix}. 
%In real-world scenarios (e.g., natural meetings~\cite{ccetin2006analysis}, conversations~\cite{barker2018chime}), only 20\% of the input audio contains overlapping speech segments.
Also, mixtures from same-gender speakers are much more difficult to separate than different-gender case~\cite{hershey2016deepCL}.
However, existing speech separation models employ a fixed processing path regardless of the input content, incurring unnecessary computation overhead.
%For example,  % LibriMix
%, e.g., natural meetings, and casual dinners. 

%This processing redundancy is especially critical for Transformer-based models; for example, unnecessarily processing a momentary silence of $0.5$ second chunk takes up $500$ additional input tokens for the Transformer computation. 

Inspired by \cite{graves2016adaptive}, we prune redundant tokens from the input sequence with the Adaptive Token Pruning technique. Specifically, each input token self-determines its redundancy and adaptively prunes itself (Figure~\ref{fig:adaptivepruning}).
%Our work is inspired by Graves' work on RNN~\cite{graves2016adaptive}. 
%Let a sequence $\{h_i^{(0)}\}_{i=1}^{T}$ and its transformation $f_T:\mathbb{R}^{H} \to \mathbb{R}^H$ (in our case, a transformer layer). 
Let an input token sequence $\{h_i^{(0)}\}_{i=1}^{T}$ of each input token $h_{i}^{(0)} \in {R}^{H}$, its transformation $f_T: {R}^{H} \to {R}^H$ (in our case, a transformer layer), and 
  a differentiable estimator 
%$f_p:\mathbb{R}^{H} \to (0,1)$ predicts the halting probability of each input token $h_{i}^{(0)} \in \mathbb{R}^{H}$. 
$f_p: {R}^{H} \to (0,1)$.
The $i$-th output token $y_i^{(N)}$ is defined,
\vspace{-0.5em}
\begin{equation}
\small
\begin{aligned}
h^{(n)}_{i} &= f_T\big(h^{(n-1)}_{i}\big|h^{(n-1)}_1,h^{(n-1)}_2, \cdots\big), \quad\text{ if } P^{(n-1)}_{i} \le P_{th} \\
p^{(n)}_{i} &= f_p(h^{(n)}_{i}),\quad\quad\qquad\qquad\qquad\quad\qquad\text{ if } P^{(n-1)}_{i} \le P_{th}\\
P^{(n)}_{i} &=
\begin{cases}
P^{(n-1)}_{i} + p^{(n)}_{i}, & \quad\mbox{if } P^{(n-1)}_{i} \le P_{th}  \\
1,  & \quad\mbox{otherwise}
\end{cases}\\
%P^{(n-1)}_{i} + p^{(n)}_{i}  \text{ if } P^{(n-1)}_{i} \le P_{th}\\
\end{aligned}
\nonumber
\end{equation}

\vspace{-0.5em}
\begin{equation}
\small
\begin{aligned}
y_i^{(n)} &=
\begin{cases}
y_i^{(n-1)} + p_{i} h^{(n)}_{i}, & \ \  \mbox{if } P^{(n-1)}_{i} \le P_{th} \\
y_i^{(n-1)} + (1- P^{(n-1)}_{i}) h^{(n)}_{i},  & \ \ \mbox{otherwise}
\end{cases}
\end{aligned}
\nonumber
\end{equation}

where $P_{th} \in [0,1]$ is the halting threshold, $P^{(0)}_{i} = 0$, and $N$ is the maximum $f_T$ iterations. So each sequence token processes $f_T$ until its halting probability $P_{i}$ reaches the threshold. This enables dynamic scaling of the transformer's \textit{width} and \textit{depth}, i.e., the number of tokens and the transformer iterations, respectively. The estimator $f_p$ has sigmoid output activation, and $f_p$ is jointly trained with the transformer $f_T$.

\noindent\textbf{Piggybacked Probability Estimator.} Prior works used a separate single-layer fully-connected network as the probability estimator $f_p$~\cite{graves2016adaptive, dehghani2018universal}. However, it incurs additional latency since it cannot be computed with the $f_T$ in parallel. Instead, we piggyback $f_p$ into the FFN of the transformer and the embedding module. Hence, the input and output of the transformer layer become a sequence of pairs of a token and its halting probability, i.e., $\{\big(h_{i}^{(n)}, p_{i}^{(n)}\big)\}_{i=1}^{T}$.

%The advantages of this design are in two folds: (i) the estimator $f_p$ can be simultaneously computed with the transform $f_T$, and (ii) $f_p$ and $f_T$ can refer to the token's halting probability. 

%\subsection{Recurrent Transformer}
\noindent\textbf{Recurrent Transformer and Time-step Encoding.} 
To reduce the model size, we employ recurrent transformer~\cite{dehghani2018universal} architecture, which iterates a single parameter-shared transformer layer instead of stacking it. 
%There are two reasons. First, every transformer layer should share the token embedding space~\cite{dehghani2018universal}, since our pruning technique weight-averages tokens from each transformer layer. 
%This reduces the number of parameters to $\frac{1}{N}$ compared to the stacked model with depth $N$. 
This reduces the number of parameters to $\frac{1}{N}$ compared to the Transformer with depth $N$. 
%Our difference from~\cite{dehghani2018universal} is how we embed the information of iteration steps. 
Opposed to \cite{dehghani2018universal} which adds the fixed time-step encoding vector to the tokens, we make the time-step information learnable by embedding it in the trainable affine parameters of the transformer's layer normalization. 
% In each step, the sequence tokens are affine-mapped to a different subspace to be uniquely processed. 
%In essence, it is equivalent to sharing only the parameters of linear layers of the transformer.
Essentially, it is equivalent to sharing only the linear layer weights of the transformer.

%\documentclass[../main.tex]{subfiles}
%\begin{document}

\begin{table*}[h!]
\centering
\fontsize{9}{11}\selectfont
 \begin{tabular}{@{\extracolsep{3pt}}c c c c c c c  c@{}}
 \toprule
 \multirow{2}{*}{Models} & \multicolumn{2}{c}{Libri2Mix~\cite{cosentino2020librimix}} & \multicolumn{2}{c}{WSJ0-2Mix~\cite{hershey2016deepCL}} & Params& \multicolumn{2}{c}{Latency (5s, 8kHz input)}  \\  
 \cline{2-3}\cline{4-5} \cline{7-8}
  & SISNRi ($\uparrow$) & SDRi ($\uparrow$) & SISNRi ($\uparrow$) & SDRi ($\uparrow$) &  (M) & CPU (s) & GPU (ms) \\  
 \midrule
 DPCL~\cite{hershey2016deepCL} & 5.9* & 6.6* & 10.8 & 11.2 & 13.6 &0.62 & 111.80 \\ 
 uPIT-LSTM~\cite{kolbaek2017multitalker} & 7.6* & 8.2* & 9.8 & 10.0 &  92.7 & 0.56 & 110.44\\
 Chimera++~\cite{wang2018alternative} & 6.3* & 7.0* & 11.5 & 11.8 & 32.9 & 0.69& 152.36 \\
 BLSTM-TasNet~\cite{luo2018tasnet} & 7.9* & 8.7* & 13.2 & 13.6 &  23.6 & 1.68 &335.58  \\
 Two-step TDCN~\cite{tzinis2020tdcn} & 12.0* & 12.5* & 16.1 & - &  8.6& 0.67& 156.11\\
 DPRNN~\cite{luo2020dualpathrnn} &14.1* & 14.6* & 18.8 & 19.0 &  2.7 & 4.36 & 207.76 \\
 Sandglasset~\cite{lam2021sandglasset} & - & - & 20.8 & 21.0 &  2.3 & 6.89&155.02\\
 DPT~\cite{chen2020dual} & 16.2 & 16.8 & 20.2 & 20.6 &  2.6 & 24.73 &336.74 \\
 Wavesplit~\cite{zeghidour2021wavesplit}& \textbf{19.5} & \textbf{20.0} & \textbf{21.0} & \textbf{21.2} &  29.0& 9.15 &274.87\\ 
 Sepformer~\cite{subakan2021sepformer}& 19.2& 19.4 & 20.4 & 20.5 &  26.0 & 3.16 &168.45  \\
 \midrule
 ConvTasNet~\cite{luo2019conv} & 12.2 & 12.7 & 15.3 & 15.6 & 5.6 & \textbf{0.40} & \textbf{34.03} \\ 
 A-FRCNN-16~\cite{hu2021afrcnn} &16.7 & 17.2 & 18.3 & 18.6 &  6.1 & 1.95&124.52\\
 A-FRCNN-16(sum)~\cite{hu2021afrcnn} &16.2 & 17.2 & 17.9 & 18.3 &  \textbf{1.7}& 1.24&118.98 \\
 %TransMask \\
 Tiny-Sepformer~\cite{luo2022tiny} &- &- &15.1 &16.1 &20.0 & 3.09&173.39\\
 Tiny-Sepformer-S~\cite{luo2022tiny} & -& -&15.2 &16.0 & 5.3 & 2.96&173.37\\
 \midrule
 \textbf{\modelname}\ &  17.2& 17.6 & 19.2& 19.5&\multirow{2}{*}{\textbf{1.47}} & \multirow{2}{*}{\textbf{1.18}}& \multirow{2}{*}{\textbf{70.88}}\\
 \textbf{\modelname\ + DM}&  \textbf{17.3}& \textbf{17.7}&\textbf{19.4} &\textbf{19.7} &\\
 \bottomrule
 \end{tabular}
\vspace{-0.8em}
 \caption{Separation performance and computational efficiency of our \textbf{\modelname}\ model compared with prior works. Latency of baselines measured from asteroid~\cite{pariente2020asteroid} toolkit with  their code~\cite{subakan2021sepformer,lam2021sandglasset,hu2021afrcnn,pariente2020asteroid} and ours~\cite{luo2022tiny}. (*) denotes results reported by~\cite{hu2021afrcnn}.
 %with existing model implementations~\cite{subakan2021sepformer,lam2021sandglasset,hu2021afrcnn,pariente2020asteroid} or our reproduction~\cite{luo2022tiny}. 
 }
 \label{tab:performance}
\vspace{-1.5em}
\end{table*}

\begin{table}[h!]
\centering
\fontsize{9}{11}\selectfont
\begin{tabular}{@{\extracolsep{0.05pt}}r l c c c c @{}}
 \toprule
 &\multirow{2}{*}{Method} &  \multicolumn{2}{c}{WSJ0-2Mix} &  Params & Latency \\  
 \cline{3-4}
  & & SISNRi($\uparrow$) & SDRi($\uparrow$) &  (M) & (ms) \\  
 \midrule
 \multicolumn{2}{c}{Sepformer} & 20.4 & 20.5 & 26.0 & 168.31 \\ 
 (1)&$+$ Recurrence& 17.1 & 17.4 & 2.25&163.67  \\
 (2)&$+$ Pruning & 17.9 & 18.1 & 2.38&95.22\\
 (3)&$+$ No Inter-T& 15.6 & 16.1 & 1.52 & 82.97\\
 (4)&$+$ AWM&19.0& 19.3 & 1.52 & 86.18\\ 
 (5)&$+$ Time Enc.&\textbf{19.4} & \textbf{19.7} & 1.54 & 86.53\\
 (6)&$+$ Piggyback& 19.3 & 19.6 & \textbf{1.47}& \textbf{80.06}\\
 \midrule
 \multicolumn{2}{c}{$K = 250$} & \textbf{19.3} & \textbf{19.6} & 1.47 & 79.94\\
 \multicolumn{2}{c}{$K = 150$} & 19.2 & 19.5 & 1.47 & 70.41\\
 \multicolumn{2}{c}{$K = 100$} & 19.1 & 19.3 & 1.47 & \textbf{67.28}\\
 \multicolumn{2}{c}{$K = 50$} & 19.0 & 19.3 & 1.47 & 68.34\\
 \midrule
 \multicolumn{2}{c}{$M = 16$} & \textbf{19.3} & \textbf{19.6} & 1.47 & 80.06\\
 \multicolumn{2}{c}{$M = 8$} & 18.6 & 18.9  & 1.47& 79.66\\
 \multicolumn{2}{c}{$M = 4$} & 18.3 & 18.6 & 1.47 & 77.42\\
 \multicolumn{2}{c}{$M = 2$} & 18.0 & 18.4 & 1.47 & 76.52\\
 \bottomrule
 \end{tabular}
\vspace{-0.8em}
 \caption{Effect of our key techniques and hyperparameters. 
 %\textit{Recurrence} setup makes the intra/inter-transformer of Sepformer recur a single transformer layer. 
 \textit{No Inter-T} setup (3) removes the inter-transformer from (2). $K$ is chunk size, and $M$ is the number of AWM slots.   }
 \label{tab:ablation}
\vspace{-2em}
\end{table}

\section{Experiment}
\label{sec:experiment}

\subsection{Experimental Setup}
\textbf{Datasets.} 
We use two datasets for evaluation: (i) WSJ0-Mix~\cite{hershey2016deepCL} and (ii) LibriMix~\cite{cosentino2020librimix} (\textit{train-360} for training, and \textit{test} for evaluation). The signal-to-noise ratio (SNR) of mixture is randomly selected in [-5,5] dB range.
%We evaluate our model's SCSS performance on the widely-used \textit{WSJ0-Mix} and \textit{LibriMix} datasets~\cite{hershey2016deepCL,cosentino2020librimix}.

%\begin{itemize}
%\noindent \textbf{$\bullet$ WSJ0-Mix~\cite{hershey2016deepCL}} is generated by mixing two or three different speakers' utterances from the Wall Street Journal corpus. The signal-to-noise ratio (SNR) is randomly selected in [-5,5] dB range. The \textit{training, validation,} and \textit{test} dataset consists of 30, 10, 5 hours speech mixtures, respectively. 
%\item 

%\noindent \textbf{$\bullet$ LibriMix~\cite{cosentino2020librimix}} 
%is a speech separation dataset in noisy environments. It 
%is composed of mixtures of two or three-speakers from the LibriSpeech corpus.
%, combined with ambient noise samples from the WHAM! dataset. %LibriMix cite
%We use \textit{train-360} ($212$ hours) for training, and \textit{test} ($11$ hours) for evaluation.
%\end{itemize}

\noindent \textbf{Metrics.} We use scale-invariant Signal-to-Noise Ratio improvements (SI-SNRi)~\cite{luo2018tasnet} and Signal-to-Distortion Ratio improvements (SDRi)~\cite{sdr} for separation accuracy.
% Should the mathematical definition of these near-standard datasets be formally introduced?
We report the number of parameters and inference latency on Intel Xeon Silver 4114 CPU and NVIDIA Titan XP GPU. 

\noindent\textbf{Training Details.} We set the AWM  size as $16$, maximum depth $16$, chunk size $150$, $8$ attention heads, embedding token size 256 and $1024$ hidden nodes of FFN. The Encoder and Decoder have kernel size $16$ and stride $8$. The model is trained for $100$ epochs on LibriMix and $200$ epochs on WSJ0 dataset, with $16$-bit mixed precision. Batch size is set to $1$. We use SI-SNR with utterance-level Permutation Invariant loss~\cite{yu2017pit}.  
We use the AdamW~\cite{loshchilov2018adamw} optimizer with $10^{-4}$ learning rate, $10^{-4}$ weight decay, and exponential learning rate decay in a factor of $0.98$.  Maximum $L_2$ norm of gradient is clipped with $1$. We also use Dynamic Mixing (DM)~\cite{zeghidour2021wavesplit} for augmentation.
%is investigated, which generates mixtures of randomly selected speaker's utterances on-the-fly. 
The entire training takes 220 hours on a Titan RTX GPU. 

%Both the convolution layer of Encoder and transposed convolution layer of Decoder uses kernel size $16$ with stride $8$.

\subsection{Comparison with Prior Works}
%We thoroughly compare our model's separation performance and efficiency with a number of existing single-channel speech separation works. 
Table~\ref{tab:performance} shows that {\modelname} is significantly more efficient than computation-intensive high-performance baselines~\cite{subakan2021sepformer, luo2020dualpathrnn,chen2020dual,lam2021sandglasset,zeghidour2021wavesplit}. {\modelname} is  $2.20\sim4.75\times$ and $2.67\sim20.95\times$ faster on GPU, CPU, respectively. 
The separation accuracy of {\modelname} is noticeably higher than computation-efficient baselines~\cite{luo2022tiny,luo2019conv,hu2021afrcnn}. The SI-SNRi of {\modelname} has increased $1.1\sim4.2$dB on the WSJ0-Mix~\cite{hershey2016deepCL} and  $0.6\sim5.1$dB on the LibriMix~\cite{cosentino2020librimix}. Interestingly,  {\modelname} is even faster than efficient baselines~\cite{luo2022tiny, hu2021afrcnn}  of $1.67\sim 2.45\times$ on GPU and $1.05\sim2.61\times$ on CPU. 
Note that Tiny-Sepformer~\cite{luo2022tiny} focuses on reducing the model size of the Sepformer~\cite{subakan2021sepformer}, and the latency rather increased. 
The model size of {\modelname} is the smallest, which is $17.7\times$ smaller than the Sepformer baseline.
Even if we scale down the transformer depth of the Sepformer to 3 or 4 with FFN size $2048$, {\modelname} is $1.26\times$, $1.65\times$ faster with $11.01\times$, and $14.59\times$ smaller model size, respectively. 
%Results show that {\modelname} is computationally efficient, lightweight, and highly performant.

%Compared models include frequency-domain approaches: DPCL, uPIT-BLSTM-ST, and Chimera++. Compared time-domain models are: BLSTM-TasNet, ConvTasNet, Two-step TDCN, Dual-path RNN, FurcaNext, DPRNN, DPT, Sandglasset, Sepformer, A-FRCNN, TransMask, Tiny-Sepformer.

% UPIT: https://github.com/JusperLee/UtterancePIT-Speech-Separation/blob/master/separate.py

\subsection{Ablation Study}
%To analyze the effect of our key technical components, we gradually apply each component  starting from the Sepformer model. 
Table~\ref{tab:ablation} shows that our techniques either contributes to efficiency ((1) $\sim$ (3), (6)) or separation accuracy ((2), (4), (5)). Note that our working memory technique greatly enhances the accuracy with minimal latency overhead of $3.8\%$ (4). Also, our adaptive token pruning noticeably cuts down the latency to $58.1\%$ while boosting the performance (2).

We also investigate the effect of two key hyperparameters: chunk size and AWM size.  Table~\ref{tab:ablation} shows that scaling down the chunk size notably decreases the latency, yet its impact on performance is marginal. 
%Surprisingly, changing the chunk size from $100$ to $50$ increases the latency. 
The reason the latency increases from $K = 100 \to 50$ is that the latency proportion of the attention mechanism  
%that the chunk size mainly affects the latency of the scaled dot-product attention mechanism (\textit{Matmul}, \textit{Matmul2} of Multi-head Attention in Figure~\ref{fig:breakdown}), and its latency proportion 
becomes insignificant with a small chunk size. In contrast, reducing the AWM size greatly diminishes the performance.  This demonstrates that our AWM plays a pivotal role in speech separation performance.

%\documentclass[../main.tex]{subfiles}
%\begin{document}
\section{Conclusion}
\label{sec:conclusion}
We proposed \modelname, an efficient and lightweight single-channel speech separation model. For computational efficiency, we substituted inter-transformer with small-sized working memory, and adaptively pruned redundant tokens from the input sequence. 
%Our core techniques are three-folds: (i) Working memory transformer, (ii) Adaptive token pruning, and (iii) Recurrent transformer. 
Experimental results on WSJ0-2Mix~\cite{hershey2016deepCL} and Libri2Mix~\cite{cosentino2020librimix} datasets demonstrate  that our model is computationally efficient yet highly performant. 
%A key characteristic of our approach is that it is orthogonal to a large number of optimization techniques on the attention mechanism, e.g., Reformer, Performer, Linformer, and Luna. We plan to integrate them into our new model and further explore their efficiency and separation performance. Since our model architecturally resembles the BERT model, we hypothesize that self-supervised pre-training will greatly enhance the separation accuracy. 

%\end{document}

\section{Acknowledgement}
This work was supported by the National Research Foundation of Korea(NRF) grant funded by the Korea government(MIST) (No. 2022R1A2C3008495).

\bibliographystyle{IEEEbib}
\bibliography{refs}

\begin{thebibliography}{10}

\bibitem{subakan2021sepformer}
C.~Subakan, M.~Ravanelli, S.~Cornell, M.~Bronzi, and J.~Zhong,
\newblock ``Attention is all you need in speech separation,''
\newblock in {\em ICASSP 2021}, 2021, pp. 21--25.

\bibitem{lutati2022sepit}
S.~Lutati, E.~Nachmani, and L.~Wolf,
\newblock ``Sepit approaching a single channel speech separation bound,''
\newblock {\em Proc. Interspeech}, 2022.

\bibitem{rixen2022qdpn}
J.~Rixen and M.~Renz,
\newblock ``Qdpn-quasi-dual-path network for single-channel speech separation,''
\newblock {\em Proc. Interspeech}, pp. 5353--5357, 2022.

\bibitem{luo2020dualpathrnn}
Y.~Luo, Z.~Chen, and T.~Yoshioka,
\newblock ``Dual-path rnn: efficient long sequence modeling for time-domain single-channel speech separation,''
\newblock in {\em ICASSP 2020}, 2020, pp. 46--50.

\bibitem{lam2021sandglasset}
M.~Lam, J.~Wang, D.~Su, and D.~Yu,
\newblock ``Sandglasset: A light multi-granularity self-attentive network for time-domain speech separation,''
\newblock in {\em ICASSP}, 2021, pp. 5759--5763.

\bibitem{cosentino2020librimix}
J.~Cosentino, M.~Pariente, S.~Cornell, A.~Deleforge, and E.~Vincent,
\newblock ``Librimix: An open-source dataset for generalizable speech separation,''
\newblock {\em arXiv preprint arXiv:2005.11262}, 2020.

\bibitem{pasternak2005working}
T.~Pasternak and M.~Greenlee,
\newblock ``Working memory in primate sensory systems,''
\newblock {\em Nature Reviews Neuroscience}, vol. 6, no. 2, pp. 97--107, 2005.

\bibitem{kumar2016brain}
S.~Kumar, S.~Joseph, P.~Gander, N.~Barascud, A.~Halpern, and T.~Griffiths,
\newblock ``A brain system for auditory working memory,''
\newblock {\em Journal of Neuroscience}, vol. 36, no. 16, pp. 4492--4505, 2016.

\bibitem{Ainslie2020ETC}
J.~Ainslie, S.~Onta{\~n}{\'o}n, C.~Alberti, V.~Cvicek, Z.~Fisher, P.~Pham, A.~Ravula, S.~K. Sanghai, Q.~Wang, and L.~Yang,
\newblock ``Etc: Encoding long and structured inputs in transformers,''
\newblock in {\em EMNLP}, 2020.

\bibitem{Zaheer2020BigBT}
M.~Zaheer, G.~Guruganesh, K.~Dubey, J.~Ainslie, C.~Alberti, S.~Onta{\~n}{\'o}n, P.~Pham, A.~Ravula, Q.~Wang, L.~Yang, and A.~Ahmed,
\newblock ``Big bird: Transformers for longer sequences,''
\newblock {\em ArXiv}, vol. abs/2007.14062, 2020.

\bibitem{graves2016adaptive}
A.~Graves,
\newblock ``Adaptive computation time for recurrent neural networks,''
\newblock {\em arXiv preprint arXiv:1603.08983}, 2016.

\bibitem{dehghani2018universal}
M.~Dehghani, S.~Gouws, O.~Vinyals, J.~Uszkoreit, and L.~Kaiser,
\newblock ``Universal transformers,''
\newblock in {\em ICLR}, 2018.

\bibitem{luo2022tiny}
J.~Luo, J.~Wang, N.~Cheng, E.~Xiao, X.~Zhang, and J.~Xiao,
\newblock ``Tiny-sepformer: A tiny time-domain transformer network for speech separation,''
\newblock {\em Proc. Interspeech}, 2022.

\bibitem{hu2021afrcnn}
X.~Hu, K.~Li, W.~Zhang, Y.~Luo, J.~Lemercier, and T.~Gerkmann,
\newblock ``Speech separation using an asynchronous fully recurrent convolutional neural network,''
\newblock {\em NeurIPS}, vol. 34, pp. 22509--22522, 2021.

\bibitem{Logeshwari2012ASO}
G.~Logeshwari and G.~S.~Anandha Mala,
\newblock ``A survey on single channel speech separation,''
\newblock 2012.

\bibitem{luo2018tasnet}
Y.~Luo and N.~Mesgarani,
\newblock ``Tasnet: time-domain audio separation network for real-time, single-channel speech separation,''
\newblock in {\em ICASSP 2018}, 2018, pp. 696--700.

\bibitem{chen2020dual}
J.~Chen, Q.~Mao, and D.~Liu,
\newblock ``Dual-path transformer network: Direct context-aware modeling for end-to-end monaural speech separation,''
\newblock {\em Proc. Interspeech}, pp. 2642--2646, 2020.

\bibitem{lin2021surveyT}
T.~Lin, Y.~Wang, X.~Liu, and X.~Qiu,
\newblock ``A survey of transformers,''
\newblock {\em arXiv preprint arXiv:2106.04554}, 2021.

\bibitem{luo2019conv}
Y.~Luo and N.~Mesgarani,
\newblock ``Conv-tasnet: Surpassing ideal time-frequency magnitude masking for speech separation,''
\newblock {\em IEEE/ACM transactions on audio, speech, and language processing}, vol. 27, no. 8, pp. 1256--1266, 2019.

\bibitem{vaswani2017attention}
A.~Vaswani, N.~Shazeer, N.~Parmar, J.~Uszkoreit, L.~Jones, A.~Gomez, L.~Kaiser, and I.~Polosukhin,
\newblock ``Attention is all you need,''
\newblock {\em NIPS}, vol. 30, 2017.

\bibitem{wu2020liteT}
Z.~Wu, Z.~Liu, J.~Lin, Y.~Lin, and S.~Han,
\newblock ``Lite transformer with long-short range attention,''
\newblock {\em ICLR}, 2020.

\bibitem{hershey2016deepCL}
J.~Hershey, Z.~Chen, J.~Le~Roux, and S.~Watanabe,
\newblock ``Deep clustering: Discriminative embeddings for segmentation and separation,''
\newblock in {\em ICASSP 2016}, 2016, pp. 31--35.

\bibitem{schulze2013wmpitch}
K.~Schulze and B.~Tillmann,
\newblock ``Working memory for pitch, timbre, and words,''
\newblock {\em Memory}, vol. 21, no. 3, pp. 377--395, 2013.

\bibitem{zhang2016auditory}
Y.~Zhang, D.~Moore, J.~Guiraud, K.~Molloy, T.~Yan, and S.~Amitay,
\newblock ``Auditory discrimination learning: Role of working memory,''
\newblock {\em PloS one}, vol. 11, no. 1, pp. e0147320, 2016.

\bibitem{kolbaek2017multitalker}
M.~Kolb{\ae}k, D.~Yu, Z.~Tan, and J.~Jensen,
\newblock ``Multitalker speech separation with utterance-level permutation invariant training of deep recurrent neural networks,''
\newblock {\em IEEE/ACM Transactions on Audio, Speech, and Language Processing}, vol. 25, no. 10, pp. 1901--1913, 2017.

\bibitem{wang2018alternative}
Z.~Wang, J.~Le~Roux, and J.~Hershey,
\newblock ``Alternative objective functions for deep clustering,''
\newblock in {\em ICASSP 2018}, 2018, pp. 686--690.

\bibitem{tzinis2020tdcn}
E.~Tzinis, S.~Venkataramani, Z.~Wang, C.~Subakan, and P.~Smaragdis,
\newblock ``Two-step sound source separation: Training on learned latent targets,''
\newblock in {\em ICASSP 2020}, 2020, pp. 31--35.

\bibitem{zeghidour2021wavesplit}
N.~Zeghidour and D.~Grangier,
\newblock ``Wavesplit: End-to-end speech separation by speaker clustering,''
\newblock {\em IEEE/ACM Transactions on Audio, Speech, and Language Processing}, vol. 29, pp. 2840--2849, 2021.

\bibitem{pariente2020asteroid}
M.~Pariente, S.~Cornell, J.~Cosentino, S.~Sivasankaran, E.~Tzinis, Je. Heitkaemper, M.~Olvera, Fabian-R. St{\"o}ter, M.~Hu, J.~Mart{\'\i}n-Do{\~n}as, et~al.,
\newblock ``Asteroid: the pytorch-based audio source separation toolkit for researchers,''
\newblock {\em Proc. Interspeech}, 2020.

\bibitem{sdr}
E.~Vincent, R.~Gribonval, and C.~Fevotte,
\newblock ``Performance measurement in blind audio source separation,''
\newblock {\em IEEE Transactions on Audio, Speech, and Language Processing}, vol. 14, no. 4, pp. 1462--1469, 2006.

\bibitem{yu2017pit}
D.~Yu, M.~Kolb{\ae}k, Z.~Tan, and J.~Jensen,
\newblock ``Permutation invariant training of deep models for speaker-independent multi-talker speech separation,''
\newblock in {\em ICASSP 2017}, 2017, pp. 241--245.

\bibitem{loshchilov2018adamw}
I.~Loshchilov and F.~Hutter,
\newblock ``Decoupled weight decay regularization,''
\newblock in {\em ICLR}, 2018.

\end{thebibliography}

\end{document}